# State-Based Modeling of Buildings and Facilities

**Univ.-Prof. Dr. Ing. M. Norbert Fisch**
IGS
TU Braunschweig
http://www.igs.bau.tu-bs.de

**Dipl.-Inform. Markus Look**
Software Engineering
RWTH Aachen University
http://www.se-rwth.de

**Dipl.-Inform. Claas Pinkernell**
synavision GmbH
http://www.synavision.de

**Dipl.-Ing. Architekt Stefan Plesser**
synavision GmbH
http://www.synavision.de

**Univ.-Prof. Dr. Bernhard Rumpe**
Software Engineering
RWTH Aachen University
http://se-rwth.de

ABSTRACT
Research on energy efficiency of today's buildings focuses on the monitoring of a building's behavior while in operation. But without a formalized description of the data measured, including their correlations and in particular the expected measurements, the full potential of the collected data can not necessarily be exploited. Who knows if a measured value is good or bad? This problem becomes more virulent as smart control systems sometimes exhibit intelligent, but unexpected behavior (e.g. starting heating at unconventional times). Therefore we defined a methodology starting already at the design of the building leading to a formalized specification of the implementation of a building's management system, which seamlessly integrates to an intelligent monitoring. DIN EN ISO 16484 proposes a method to describe functional requirements in an easy to understand way. We extended its use of state machines to our proposed concept of state based modeling. This proved to be a wholesome approach to easily model buildings and facilities according to the DIN EN ISO 16484 while providing the possibility to apply sophisticated and meaningful analysis methods during monitoring.

## 1. INTRODUCTION

Saving **energy in today's world of rising $CO_2$** emission is an inevitable task. Most current tools focus on monitoring the energy efficiency of buildings and facilities. But saving energy and natural resources starts in the earlier phases of every **building's develop**ment process and depends on the overall requirements and implementation quality.

With our already published concept of the Energy Navigator (Fisch et al., 2010) we proposed a methodology that addresses the qualitative aspects in an early phase of this process. Figure 1 shows our closed optimization circuit starting at the design phase. We provide a convenient tool to specify the desired behavior of the building management system (BMS) as formalized requirements documentation. This formal specification can be implemented directly, avoiding errors which might otherwise be detected in later phases (integrated quality assurance). The implementation of such a specification into a PLC controller can be done e.g. using existing programming tools (Siemens, 2004). To check the implementation, analysis algorithms can be derived and automatically processed from the specification to proof the quality of service during operation. The monitoring process is supported by target-aimed data preparation and visualization. The collected information can easily be used in the **optimization phase to improve the building's** efficiency by adapting the specification and implementation (feedback loops). By introducing such a methodology one omits inconsistencies between the planned functional requirements of a BMS and the actual realization.

To realize functional requirements the DIN EN ISO 16484 proposes several concepts to describe the functional behavior of facilities. Our state based modeling approach combines the concept of state machines from the area of computer science and the concepts proposed in the DIN EN ISO 16484. By describing such a functional requirement in an easy to understand way, we show that we are able to specify and check functional requirements fully automatic. This provides an extension to the already existing concepts of the Energy Navigator like metrics, rules, functions, characteristics and other elements. Using the information provided by the BMS about the state of a building or a facility we show how to model all functional requirements in a state-based way combining it with the aforementioned concepts to specify requirements for

```
rule isNightMode {
    sensors {
        I1 = "000-000-001";
    }
    if isNight
    then I1 = 18.0
    else true
}
```

Listing 1: Textual representation of a specified rule

every possible state. By connecting the existing concepts and the state-based modeling approach we provide a new methodology for planning and monitoring functional behavior. With this approach we present new options to specify more fine-grained quality criteria in the design-, monitoring- and optimization phase.

The paper is structured as follows: First we explain the concept of the Energy Navigator and motivate a concrete example for using our approach. After that we introduce our concept of state based modeling. We provide a theoretical foundation and show the adaptations we made to the existing concept of state machines, necessary to use it together with a building automation process. Subsequently we show how this approach can be extremely helpful during the whole process of implementing a building automation system together with visual evaluation means to gather data that enables the user to easily see errors in the automation system. After that we conclude our approach with a discussion of the benefits and open points for future work.

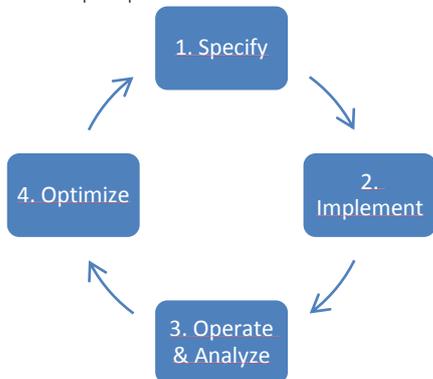

Figure 1: Closed optimization circuit starting with a formal specification

## 2. BASIC PRINCIPLES

The Energy Navigator is a technical state of the art software platform, using client- server mechanisms, providing the possibility to not only monitor but also specify the behavior of a BMS. The platform provides all means necessary to get a closed circuit between specification, monitoring, data analyses and optimization. Monitoring of buildings and providing sophisticated analyses means handling mass data. The platform is able to import data either automatically, by using OPC (Mahnke et al., 2009) to directly read values from the BMS, or manually, from a data logger, in discrete intervals. Since we created a highly adaptable platform we are able to scale up the data collection to one data point each second for a single sensor. But under normal conditions a resolution of collecting one data point every 15 minutes for a single sensor has proven to be feasible and state of the art in building automation systems. Even this coarse grained data collection leads to a lot of information that needs to be stored and processed by the platform. Consider a building having 1000 sensors each producing a data point every 15 minutes. Thus we get 96000 data points a day from a single building which result in about 35 million data points a year. To keep the performance and scalability we use cloud computing (Lenk et al. 2009, Zhang et al. 2010) based storage techniques. Additionally most modern building automation systems provide the possibility to not only log data points but to log also markers signaling different modes a building or part of it resides in.

Each data collection is followed by a preprocessing step ensuring that all the values have equidistant timestamps and have passed outlier detection. Since the building automation system cannot ensure this, might miss single values or collect them a bit too late we use a multistep algorithm to calculate the correct timestamps for each data point or can even interpolate single data points. We will later see that our analysis and our concept of state based modeling are affected by this. Having collected all the data the user needs to be able to create custom analyses for the monitored building. To support these tasks the Energy Navigator platform provides the possibility to create several elements aiding the analyses of mass data measured in a building. We therefore created a Domain Specific Language (Karsai et al., 2009) for specifying and modeling buildings. To create such a language we use our framework MontiCore (Krahn et al., 2010). This Language consists of several elements describing the tools at hand for an energy expert. These elements are: metrics, characteristics, time routines, constants, functions and rules. We explain the concept of functions and rules in more detail in section 2.1 whereas the other concepts are explained in (Fisch et al., 2010). To provide aggregated information over the specified analyses we created different plots, like a standard line plot, a scatter plot or a carpet plot displaying data aggregated over time. Since we decoupled the user interface from the actual server backend computing the analyses, our platform is able to be used from different kinds of user devices, like personal computers running the expert tool, tablets or even smartphones to get only necessary information relevant to the user on site.

```
rule arePeoplePresent {
    sensors {
            I1 = "000-000-002";
    }
    I1 > 0
}
```
Listing 2: The *arePeoplePresent* rule

```
rule referenceValueChange {
    sensors {
            I1 = "000-000-003";
    }
    (I1 >= referenceValue - 3)
      or
    (I1 <= referenceValue + 3)
}
```
Listing 3: The *referenceValueChange* rule

### 2.1 The Rule Domain Specific Language

This section provides a detailed description of the Domain Specific Language (DSL) we created to formulate rules or functions. We focus on the concept of rules and omit the functions since the basic difference between them is that functions always evaluate to a numeric value while rules evaluate to a logical (true/false) value. Listing 1 shows a textual representation of a specified rule. We explicitly present this rule from a technical point of view. Our tool that supports the creation of such rules abstracts from the concrete syntax and provides the user with an intuitive graphical user interface to create them. This rule is named *isNightMode* and can be referenced as such from other elements or other documentation. The next block assigns a readable identifier to an existing sensor which is known by its BMS-ID inside the building management system. We can use this identifier throughout the rule to reference the sensor. The rule's body contains the actual specification. The example specifies a comparison between the sensor and a reference value. To fully understand the rule one has to know the possibility of cross referencing other elements, like the time routine called *isNight*. Keeping in mind that we use equidistant time steps to evaluate our elements one has to think of such a time routine as an evaluation to true or false for any given timestamps. One could specify the time routine as "*every day between 10 p.m. and 6 a.m. the building should be in night mode*". So for every timestamp inside the specified range this evaluates to true or false otherwise. To sum it up, the rule specifies, that the sensor which collects the data important to us, should measure the value **18.0** each time the building is in night mode. If **it isn't in night mode we don't care in this context** and evaluate to **true**. After evaluating this using a carpet plot we can easily see if there are any occurrences where our specification may not hold for the given time range by using a carpet plot. Beside the if/then/else and the equals operator (=) we provide a comprehensive set of operators like && (AND), || (OR), implies, ! (NOT), comparisons (>, <, <=, >=), arithmetical operators like +, -, *, / and library functions like the absolute function. Furthermore the language can handle parenthesized operations. These operators can be nested unrestricted. Of course we are able to check if there are errors or an unreasonable nesting of the operators in place and provide the user with some feedback.

### 2.2 Concrete Example

To clarify our approach we give a concrete example of how to apply our approach of state based modeling in combination with the existing concepts to the specification and monitoring of a building.

Consider a basic room temperature control system as specified in (Arbeitskreis der Professoren für Regelungstechnik, 2004). It consists of several elements like a radiator, a chilling system, a window contact, a shutter control, a temperature control interface and a presence sensor. Additionally it has four modes of operation, namely main mode, sleep mode, night mode and antifreeze mode. The main mode is characterized by people's presence being able to alter the reference value about ±3K. **The sleep mode is used to reach the main mode's reference** value fast if nobody is present. Additionally fig. 2 provides a characteristic for the temperature control being in either main mode or sleep mode. In contrast to the main mode the night mode is not characterized by the presence of people but by a timing mechanism. During this mode the temperature is kept low and all elements are either shut off or kept on a minimal level. Additionally the antifreeze mode may shut down the heating or chilling system if a window is opened and a minimal reference threshold is not undercut. During the remaining parts we focus on the main mode and its conditions. The conditions belonging to the other modes can be specified analogously. One can identify the following conditions for being in the main mode as

- *people have to be present*
- *the reference value can be changed by ±3K at most*
- *the control needs to react as specified in the according characteristic*

Apart from the characteristics the conditions are specified in an informal way. But these informal requirements are used by the control engineer to implement the room temperature control system leaving room for misunderstandings and interpretations which may lead to erroneous implementations. Beside failures during the implementation there is no possibility to efficiently monitor the specifications. This shows the earlier described discrepancy between planning, realization and monitoring. Using the concepts the Energy

Navigator provides we can formulate these informal requirements in a more formal way. First, we specify a rule called *arePeoplePresent* to derive logical values from the presence sensor. The corresponding rule is shown in listing 2. Furthermore we need to specify another rule called *referenceValueChange* shown in listing 3. As this rule is quite complex we want to explain it in more detail. First of all we can see that we use the identifier *I1* for referencing the sensor that actually measures the value that people in the building configured. Furthermore we can see that the most complex part is specified in the if-condition. The condition consists of a disjunction connecting two subconditions. The former specifying that the actual measured value is not allowed to be smaller than the original *referenceValue* subtracted by three. The latter specifies exactly the opposite. Thus the rule is always evaluated to true if and only if the **actual value doesn't deviate from the reference** value by at most three. The used *referenceValue* inside the rule is itself a reference to a configurable value that is defined elsewhere. As we have now modeled the first two requirements we have to model the last informal requirement. For this purpose we use a characteristic defining the behavior. The characteristic opposes two sensors that measure according to our example an action and the reaction of the temperature control system. Figure 2 shows the graphical representation of such a characteristic.

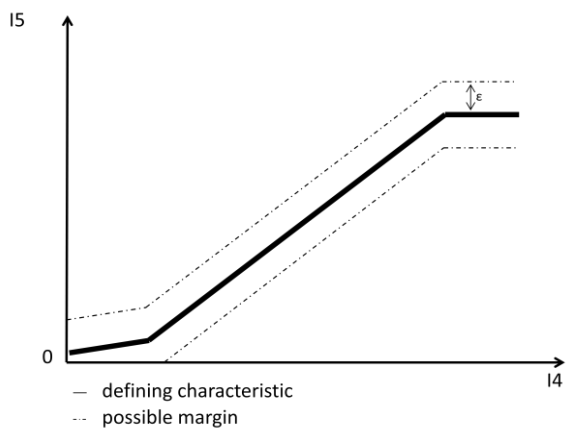

Figure 2: A characteristic defining the reaction of the temperature control system

The solid line represents the defining characteristic while the dashed line defines a possible margin due to the reaction of the system. Since it can only react it cannot follow the defining characteristic exactly. Using the example and the previous informal requirements we have shown how to derive a formal specification from the requirements. As the example **doesn't cover the fact t**hat a building or a facility can reside in different modes we expand our concept with certain elements to also cover this in a formal specifying way.

In the next section we introduce our approach of state based modeling and show how to model and how to evaluate the specification leading to a wholesome approach for planning, implementation and monitoring.

## 3. STATE MODELS

State machines (Harel, 1987) and especially hierarchical state machines are a common concept in computer science to describe the behavior (Rumpe, 1996) of a system. A state machine model consists of states and transitions. There are a lot of variants of state machines, e.g. accepting and recognizing machines, Moore (Moore, 1956) and Mealy (Mealy, 1955) machines or UML statecharts (Rumpe, 2004).

The variants define additional elements like stimuli (actions or conditions to switch from one state to another), outputs, conditions and invariants. Usually states can be tagged as start or final states. In some definitions states can be hierarchically decomposed, which means, that states may contain substates. This eases understanding of complex decomposable state machines since states can be viewed as black boxes at the top level and can be decomposed by inspecting them.

First we introduce a simple state machine definition consisting of states, transitions and stimuli. An adaption of this model for the use in facility planning is presented in the next subsection.

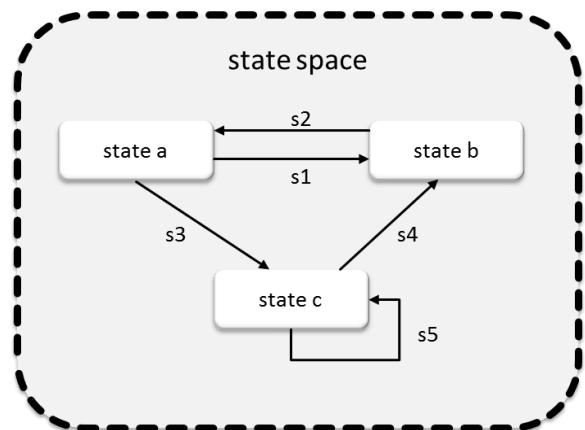

Figure 3: An abstract state machine model

Figure 3 shows three states *a*, *b* and *c*. There are transitions between the states annotated with stimuli *s1...s5*. As an example the state machine can switch from state *a* to state *b* if stimulus *s1* occurs in the system. Another notation for this behavior is shown in table 1. As you can see the graphical state machine notation is much more comprehensive.

| State / Stimulus | state a | state b | state c |
|---|---|---|---|
| s1 | state b | - | - |
| s2 | - | state a | - |
| s3 | state c | - | - |
| s4 | - | - | state b |

Table 1: state transition table

### 3.1 State-based Modeling

Picking up the previous example of the room temperature control system we propose a state based view on the description of the system. Figure 4 shows a graphical notation of the formulated modes and their conditions. The figure shows several aspects. On the one hand we can see the different elements needed to model the different modes. These elements are referenced by their name as explained in the description of listing 1. We need a sensor measuring the values that are compared to the reference value. Additionally we need two more sensors, one being a presence sensor that stores information about whether persons are present at the measured timestamp and one sensor storing the information whether the window was opened. Apart from that we make use of a time routine element to specify the interval considered as night and two characteristics specifying the reaction of the temperature control in different modes. That way we get a formalized model of each mode to be used in further analyses.

The transitions in figure 4 show possible changes from one mode to another. In contrast to figure 3 we do not model any stimuli at transitions. But they can be modeled to give the control engineer a specification on how to implement the BMS.

Transferring our example to the idea of state based modeling and to provide a formal definition we show the similarities between the well-known concept of state machines from computer science and our proposed methodology. We can consider the example model to consist of different states and transitions where a state describes the mode of the facility and the transitions describe the conditions that must be fulfilled to change the state. As we can see transitions are not required to have a condition. These transitions are triggered spontaneously thus switching the state of the system. This shows the first simplification of our model regarding the original model of a state machine. Furthermore transitions are directed and allow switching from one state to another but not vice versa in state machines. Transitions may also be loops remaining in the same state when triggered. In the automated evaluation of our model we consider undirected transitions that allow switching between both associated states and we omit the possible occurrence of loops. In subsection 3.3 we explain in more detail that it is actually not possible for the automated evaluation to have conditions for transitions and to use **loops. Additionally we don't** model an explicit initial and final state since we specify a continuously running system where this information is not needed. Nevertheless one can surely use these elements in the specification phase to communicate among the stakeholders. Thus **we don't restrict the user to use these elements but don't regard** them during the automated evaluation. These simplifications lead to a rather compact basic definition of a state space. Both states and transitions between the states define a state space:

$$SS = (S, T)$$
$$S = \{s_1, \cdots, s_n\}$$
$$T = \{(s_i, s_j)\}, 1 \leq i, j \leq |S|$$

Where S is the set of all states, T is the set of all possible transitions between all states and SS the state space defined by its states. A state can contain several sub elements as we motivated in our example. These elements belong to our DSL for specifying and modeling a building and can be evaluated to either a numerical value or a logical value. We abstract from defining each element formally since the important aspects are the

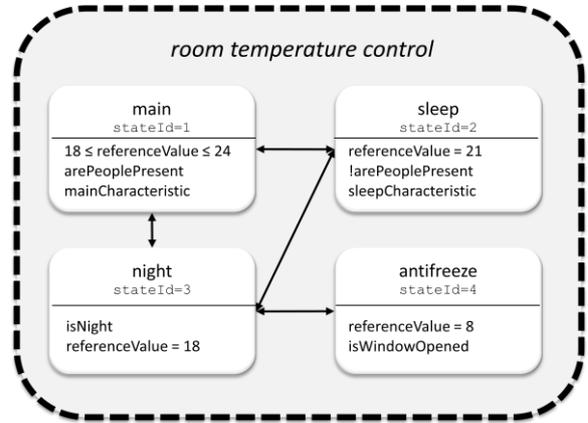

Figure 4: A modeled state space with its states

evaluation and the referencing concept that can be used inside a state. Thus a state is defined as a tuple:

$$s = (id, E).$$

In this case E should define a set of elements belonging to our language. Furthermore we omit the name of a state and a state space in the definition since it is only used for documentation issues. As mentioned earlier modern building management systems are able to monitor markers. The markers to be monitored are specified beforehand and need to be

```
rule isMainMode {
    arePeoplePresent and
    referenceValueChange and
    satisfiesCharacteristic
}
```
Listing 4: An integrating rule for checking the satisfaction of a state

```
rule isStatespaceSatisfied {
    isMainMode or
    isNightMode or
    isSleepMode or
    isAntifreezeMode
}
```
Listing 5: An integrating rule for checking the satisfaction of a state space

implemented in the BMS. During the specification phase the energy expert can define the state space including the needed states representing the different modes a building or a facility can reside in. By doing so he directly specifies the different discrete values for the markers that have to be monitored by the BMS. This information can be used by the control engineer to implement the BMS. During the monitoring phase we are able to use the information provided by the markers to enable the automated analyses since we can monitor the mode of the BMS and analyze it against the given specification.

We explicitly want to point out the important role of rules inside the concept of states. Since the other elements are used to get helpful methods in defining concepts regarding the building to be specified, the rules connect the elements and specify the constraints on a building. Due to the referencing concept we can use every element inside a rule and get a logical value as the result. This also means that we sometimes need to encapsulate other elements inside rules to get that logical value. Regarding our example we need to create a rule stating that it should evaluate to true if the reference value lies inbetween the given margin of the defining characteristic or false otherwise.

### 3.2 Evaluation Semantics

After specifying a facility and implementing the BMS we need to be able to evaluate the modeled state spaces. Therefore we define the semantics of creating such a model with the included elements. First we define the semantics of each state consisting of several elements. A state $s_i$ can either be satisfied at a given timestamp or not. For being satisfied the

$$s_i = \bigwedge_j r_j, \; r_j \in R, 1 \leq j \leq n$$

characterizes the state where R represents the set of contained rules in a state $s_i$. So every single rule has to be satisfied at a given time for the state to be satisfied as well. For the evaluation of a single state space we have to consider all included states. Therefore the following has to hold:

$$\exists_i \forall_j : s_i \wedge (s_j \rightarrow i = j), s \in S, 1 \leq i,j \leq |S|$$

This means that exactly one state may be satisfied at a given time. If there exist a second state that is satisfied then this state has to be the same state as the already satisfied one. If more than one state is allowed to be satisfied at the same time we have to soften this restriction.

$$\exists_i : s_i, s \in S, 1 \leq i \leq |S|$$

The Energy Navigator can be configured to use either possibility. This enables the user to decide if the building or facility can be in more than one state at the same time or not. The fact that only one state may hold at a given time enables us to efficiently monitor the provided marker representing the actual state of a facility. In the formal description of the state space we left out the fact that the state space itself may also contain rules which have to be satisfied completely in addition to the one single satisfied state. Picking up our example of specifying the main mode of a temperature control system we already have shown two rules describing the informal requirements. We encapsulate the defining characteristic inside another rule called *satisfiesCharacteristic*. This rule simply checks if each measured data point for a given timestamp does at most deviate by a specified ε and is defined analogous to listing 2 and 3. By defining these rules we are now able to apply our evaluation semantics.

### 3.3 Unmonitored Events

Our definition of a state space to be analyzed automatically relies on having the possibility to use a transition from each state to any other state and it relies on not having conditions specified at transitions. We want to reason about this decision since it seems to be a constraint on the model. Keeping in mind that the Energy Navigator is not a real time application and therefore doesn't collect live data one can see that we are only able to look at discrete snapshots of the monitored building. In the scenario where we collect quarter-hourly data points we might miss some events. Thus we only know in which state the building resides at the given snapshot but cannot monitor if we were in several different states in the last 15 minutes as well as we are not able to monitor if conditions are fulfilled when using a transition since we can't monitor for sure the actual point in time where the transition is used. The same holds for omitting loops which cannot be monitored. Furthermore, since we start at a given point in time to monitor a running system we get the first state the building resided in since monitoring. But this is not necessarily an initial state and we therefore don't consider initial or final states. Nevertheless this doesn't hinder the user to use loops or to use conditions for transitions during the specification phase. The user can even omit some transitions if they are irrelevant from the specification point of view. One can use it for communicating and providing only the information needed by other stakeholders.

4. STATE SPACE EVALUATION

Since we already defined all the elements necessary to specify the main mode we now combine them into a meaningful analysis. From the rules nested in a state we can automatically infer another rule combining the other rules. This is shown in Listing 4. This way we compute for any given timestamp the information if a single state is satisfied. We can use this information directly by using a carpet plot shown in figure 5.

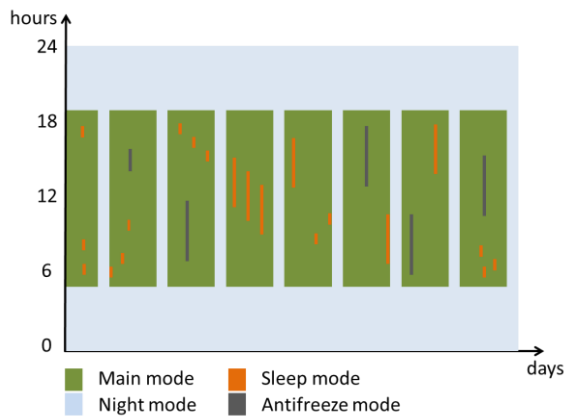

Figure 5: A carpet plot showing which state is satisfied at which point in time.

One can see in an aggregated fashion which state is satisfied at which point in time and can check it against the monitored marker values. This analysis helps at easily detecting wrong states since we would directly be able to see if the night mode was satisfied during working hours or if the temperature control system would constantly stay in main mode. Furthermore we can compare these values automatically against the monitored marker values. We then can easily derive discrepancies between the mode the BMS resides in and the actual monitored mode.

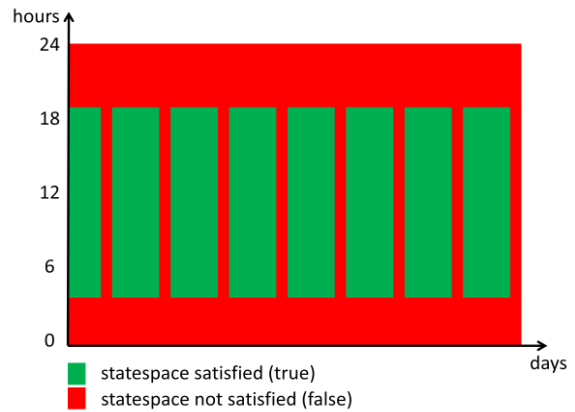

Figure 6: A carpet plot showing at what points of time the state space is satisfied or not satisfied

Apart from providing the information which state is satisfied we can also provide information if the state space as such is satisfied. This enables us to check if there exist use cases not modeled or the BMS behaves completely different to the specification. As we can see in figure 6 there are several red points that display that the complete building is in an undefined state. As an example we can see that the state space is not satisfied after working hours. This may lead to the conclusion that there has to be an error in the implementation which can now efficiently be solved.

5. THE ENERGY NAVIGATOR

The Energy Navigator subsumes capabilities for planning, monitoring and analysis purposes for a quality assured building lifecycle. The target groups of the software are building owners, facility managers and engineering consultants.

The Energy Navigator software is developed as a product in a cooperation of the Institute of Building Services and Energy Design at TU Braunschweig University[1], the Software Engineering Department at RWTH Aachen University[2] and the synavision GmbH[3] Aachen. At the end of 2011 a pilot phase will start where practical use will be evaluated. The Energy Navigator will be available as a product in 2012. The software can on the one hand be used for planning, monitoring and analyses.

---

[1] Institute of Building Services and Energy Design, TU Braunschweig University. http://www.igs.bau.tu-bs.de/
[2] Software Engineering, RWTH Aachen University. http://www.se-rwth.de
[3] synavision GmbH. http://www.synavision.de

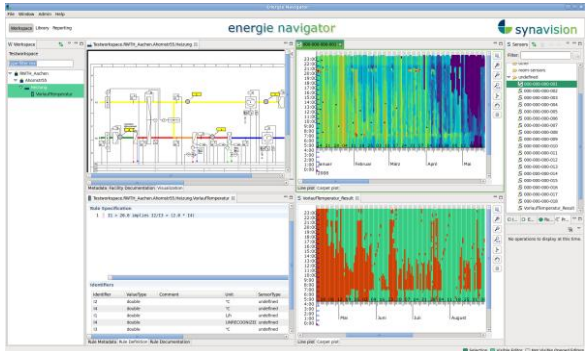

Figure 7: Screenshot of the Energy Navigator software

On the other hand the Energy Navigator can be used as a platform for a lot of technical mesh-ups, e.g. reporting platforms, information monitors and customized websites.

## 6. CONCLUSION

We have shown our concept of state based modeling throughout the whole development process of a building. We first provided the means necessary to specify and document the supposed behavior of a building and showed the analyses methods applicable during the monitoring phase. While the DIN EN ISO 16484 encourages to model functional requirements analogous to our supposed modeling approach we have shown a concrete implementation and a platform that supports this. Once we have created such a specification we can use it to communicate it throughout all succeeding phases. The electrical engineer can make use of it to implement the PLC control and the energy expert can use it to efficiently detect errors where energy is wasted in the building. We have shown the connection to the well known concept of state machines that are widely researched in computer science and provided a formal definition of our concept of state based modeling. In our opinion using this method can help to detect incomplete specifications or errors very early and therefore can help to reduce expenses and $CO_2$ emission. It would be very helpful for the monitoring and the analysis if BMS were able to monitor the exact events when changing modes. We could use this information to further improve the analysis by monitoring real changes and state changes and would not need to rely on snapshots. We then could even monitor transitions with conditions or not specified transitions that are used by the BMS. Of course these **problems wouldn't occur if the monitoring sys**tem would monitor in real time. But as mentioned in the beginning this would lead to an immense amount of collected data making it infeasible to compute analysis in an appropriate amount of time.

## 7. REFERENCES

Arbeitskreis der Professoren für Regelungstechnik, 2004**.** *Digitale Gebäudeautomation*. 3. vollständig überarbeitete und ergänzte Auflage. Berlin : Springer-Verlag.

DIN EN ISO 16484-3, Issue 2005-12

Fisch, N.M., Kurpick, T., Pinkernell, C., Plesser, S. and Rumpe, B. 2010. *The Energy Navigator - A Web based Platform for functional Quality Mangement in Buildings*. Kuwait City, Kuwait : s.n., October. Bd. Proceedings of the 10th International Conference for Enhanced Building Operations (ICEBO' 10).

Harel, D. 1987. *A Visual Formalism for Complex Systems*. *Science of Computer Programming*: 231–274.

Karsai, G., Krahn, H., Pinkernell, C., Rumpe, B., Schindler, M. and Völkel, S. 2009. *Design Guidelines for Domain Specific Languages*. In: Proceedings of the 9th OOPSLA Workshop on Domain-Specific Modeling (DSM' 09) Helsinki School of Economics. TR no B-108. Orlando, Florida, USA.

Krahn, H., Rumpe, B. and Völkel, S. 2010. *MontiCore: a Framework for Compositional Development of Domain Specific Languages*. In: International Journal on Software Tools for Technology Transfer (STTT), Volume 12, Issue 5, pp. 353-372.

Lenk, A., Klems, M., Nimis, J., Tai, S. and Sandholm, T. 2009. What's Inside The Cloud? An Architectural Map of Cloud Landscape. *CLOUD'09*, May 23, Vancouver, Canada.

Mahnke, W., Leitner, S.-H., Damm, M. 2009. OPC Unified Architecture. Springer Verlag.

Mealy, G.H. 1955**.** *A Method to Synthesizing Sequential Circuits*. In: Bell Systems Technical Journal: 1045–1079.

Moore, E.F. 1956. *Gedanken-experiments on Sequential Machines*. In: Automata Studies, Annuals of Mathematical Studies (Princeton, N.J.: Princeton University Press) (34): 129–153.

Rumpe, B. 2004. *Modellierung mit UML*. Springer.

Rumpe, B. 1996. Formale Methodik des Entwurfs verteilter objektorientierter Systeme. Herbert Utz Verlag Wissenschaft, ISBN 3-89675-149-2.
Dissertation Technische Universität München.

Siemens 2004. SIMATICS7-HiGraph V5.3 Programming State Graphs Programming and Operating Manual. 10[th] Edition.

UML. The Unified Modeling Language (UML) specification.
http://www.omg.org/spec/UML/2.3/Infrastructure/PDF/

Zhang, Q., Cheng, L., Boutaba, R. 2010.Cloud computing: state-of-the-art and research challenges. In: Journal of Internet Services and Applications(1): 7-18. Springer-Verlag